\documentclass[preprint,showpacs,preprintnumbers,amsmath,amssymb]{revtex4}

\usepackage{graphicx}

\begin{document}


\title{
Thouless-Anderson-Palmer Approach for Lossy Compression
}

\author{Tatsuto Murayama}
\affiliation{%
RIKEN Brain Science Institute, Hirosawa 2-1, Wako, Saitama 351-0198, Japan
}%


\date{\today}

\begin{abstract}
We study an ill-posed linear inverse problem,
where a binary sequence will be reproduced using a sparce matrix.
According to the previous study,
this model can theoretically provide an optimal compression scheme for
an arbitrary distortion level, though the encoding procedure
remains an NP-complete problem.
In this paper, we focus on the consistency condition for a dynamics model
of Markov-type to derive an iterative algorithm,
following the steps of Thouless-Anderson-Palmer's.
Numerical results show that the algorithm can empirically saturate
the theoretical limit for the sparse construction of our codes,
which also is very close to the rate-distortion function.
\end{abstract}

\maketitle

Lossy compression is quite important in our modern life.
One first encodes information into an appropriate form,
which will be decoded to reproduce similar sequence.
The theoretical framework for this kind of compression scheme
with a fidelity criterion is called the rate-distortion theory,
which consists an important part of the Shannon's information theory~\cite{art:shannon}.

We start by defining the concepts of the rate-distortion
theory and stating the simplest version of the main result~\cite{book:cover-thomas}. 
Let $J$ be a discrete random variable with alphabet $\mathcal{J}$. 
Assume that we have a source that produces a sequence
$J_1,J_2,\cdots,J_M$, where each symbol is randomly drawn from a
distribution. 
We will assume that the alphabet is finit.
Throughout this paper, we use vector notation to represent
sequences for convenience of explanation:
$\boldsymbol{J}=(J_1,J_2,\cdots,J_M)^T \in \mathcal{J}^M$.
Here,  
the encoder describes the source sequence 
$\boldsymbol{J} \in \mathcal{J}^M$ by a codeword 
$\boldsymbol{\xi} = f(\boldsymbol{J}) \in \mathcal{X}^N$. The decoder represent
$\boldsymbol{J}$ by an estimate $\boldsymbol{\hat J}=g(\boldsymbol{\xi})\in
\hat{\mathcal{J}}^M$. 
Note that $M$ represents the length of a source
sequence, while $N$ represents the length of a codeword. 
In this case, the rate is defined by $R=N/M$.
Note that the relation $N < M$ always holds when a compression is considered; therefore, $R<1$ also holds. 

A distortion function is a mapping
 $d:\mathcal{J}\times \hat{\mathcal{J}} \to \boldsymbol{R}^{+}$ 
 from the set of source alphabet-reproduction alphabet pairs into the set
 of non-negative real numbers.
Intuitively, the distortion $d(J,\hat{J})$ is a measure of the cost of
representing the symbol $J$ by the symbol $\hat{J}$. This definition is
quite general. In most cases, however, the reproduction alphabet
$\hat{\mathcal{J}}$ is the same as the source alphabet
$\mathcal{J}$. Hereafter, we set $\hat{\mathcal{J}}=\mathcal{J}$ and the
following distortion measure is adopted as the fidelity criterion; 
the Hamming distortion is given by
\begin{align}
  d(J,\hat{J})=
   \begin{cases}
    0 & \mathrm{if\ } J = \hat{J} \\
    1 & \mathrm{if\ } J \neq \hat{J}
   \end{cases} \ ,
\end{align}
which results in a probable error distortion, since the relation 
$E_J[d(J,\hat{J})]=\mathcal{P}[J \neq \hat{J}]$ holds, where 
$E_J[\cdot]$ represents the expectation and $\mathcal{P}[\cdot]$ the
probability of its argument. The distortion measure is so far defined on
a symbol-by-symbol basis. We extend the definition to sequences. 
The distortion between sequences 
 $\boldsymbol{J}, \boldsymbol{\hat J} \in \mathcal{J}^M$ is  
 defined by 
 $d(\boldsymbol{J},\boldsymbol{\hat J})
=(1/M)\sum_{\mu=1}^M d(J_\mu,\hat{J}_\mu)$. 
Therefore, the distortion for a sequence is the average distortion per symbol
of the elements of the sequence. The distortion associated with the code is defined as 
$D=E_{\boldsymbol{J}}[d(\boldsymbol{J},\boldsymbol{\hat J})]$, 
where the expectation is with respect to the probability distribution on
 $\mathcal{J}$. 
A rate-distortion pair $(R,D)$ should be achievable if
a sequence of rate-distortion codes $(f,g)$ exist with
$E_{\boldsymbol{J}}[d(\boldsymbol{J},\boldsymbol{\hat J})]
\le D$ in the limit $N \to
\infty$. Moreover, the closure of the set of achievable rate-distortion
pairs is called the rate-distortion region for a
source. Finally, we can define a function to describe the boundary;
 the rate-distortion function $R(D)$ is the infimum of rates
 $R$, so that $(R,D)$ is in the rate-distortion region of the source
 for a given distortion $D$.

In this paper, we restrict ourselves to a
binary source $\boldsymbol{J}$ with a Hamming distortion measure for simplicity. We assume that
binary alphabets are drawn randomly, i.e., the source is not biased to rule out the possiblity of compression due to redundancy. We
now find the description rate $R(D)$ required to describe the source
with an expected proportion of errors less than or equal to
$D$. In this simplified case, according to Shannon, the boundary can be
written as follows; 
 the rate-distortion function for a binary source with Hamming
 distortion is given by
 \begin{align}
  R(D)=
   \begin{cases}
    1-h_2(D) & 0 \le D \le \frac{1}{2} \\
    0 & \frac{1}{2} < D
   \end{cases}, \label{eq:2}
 \end{align}
 where $h_2(\cdot)$ represents the binary entropy function.

Next we introduce a toy model for the lossy compression. We use
the inverse problem of Sourlas-type decoding to realize the optimal
encoding scheme, a variation of which has recently been investigated by some
information theorists~\cite{proc:matsunaga-yamamoto}. 
As in the previous paragraphs, we assume that binary alphabets are drawn
randomly from a non-biased source and that the Hamming distortion
measure is selected for the fidelity criterion. 
Theoretically speaking, it has been reported that the typical distortion
could be well captured by the Parisi one-step RSB scheme,
giving the physical interpretaion of $R(D$)~\cite{proc:murayama-okada:nips}.
In this paper, we will discuss an actual encoding technique for this optimal
family of codes.

Firstly,
we take the Boolean representation of the binary alphabet $\mathcal{J}$,
i.e., we set $\mathcal{J}=\{0,1\}$. We also set $\mathcal{X}=\{0,1\}$ to
represent the codewords.
Let $\boldsymbol{J}$ be an $M$-bit source sequence,
$\boldsymbol{\xi}$ an $N$-bit codeword, and $\boldsymbol{\hat J}$ an
$M$-bit reproduction sequence. 
Here, the encoding problem can be written as follows.
Given a distortion $D$ and a randomly-constructed Boolean matrix $A$ of
dimensionality $M \times N$, we find the $N$-bit codeword sequence
$\boldsymbol{\xi}$, which satisfies
\begin{align}
   \boldsymbol{\hat J}&=A\boldsymbol{\xi} \quad \pmod{2} \ , \label{eq:encoding}
\end{align}
where the fidelity criterion
$D=E[d(\boldsymbol{J},\boldsymbol{\hat J})]$
holds, according to every $M$-bit source sequence $\boldsymbol{J}$. 
Note that we applied modulo $2$ arithmetics for the additive
operations in (\ref{eq:encoding}). 
In our framework, decoding will just be a linear mapping 
$\boldsymbol{\hat J}=A\boldsymbol{\xi}$, while encoding remains
an NP-complete problem.

Let the Boolean
matrix $A$ be characterized by $K$ ones per row and $C$ per column~\cite{art:kabashima-saad:replica}. 
The finite, and usually small, numbers $K$ and $C$ define a
particular code. The rate of our codes can be set to an arbitrary value
by selecting the combination of $K$ and $C$. We also use $K$ and $C$ as
control parameters to define the rate $R=K/C$. If the value of
$K$ is small, i.e., the relation $K \ll N$ holds, the Boolean matrix $A$
results in a very sparse matrix. 
By contrast, when we consider densely constructed cases, 
$K$ must be extensively big and have a value of $\mathcal{O}(N)$. 
We can also assume that $K$ is not $\mathcal{O}(1)$ but $K \ll N$
holds. 

The similarity between codes of this type and Ising spin systems was
first pointed out by Sourlas, who formulated the mapping of a code onto
an Ising spin system Hamiltonian in the context of error-correcting
codes~\cite{art:sourlas:nature}. 
To facilitate the current investigation, we first map the problem to that
of an Ising model with finite connectivity following
Sourlas' method. We use the Ising representation
$\{1,-1\}$ of the alphabet $\mathcal{J}$ and
$\mathcal{X}$ rather than the Boolean one $\{0,1 \}$; the elements of
the source $\boldsymbol{J}$ and the codeword sequences
$\boldsymbol{\xi}$ are rewritten in Ising values,
and the reproduction
sequence $\boldsymbol{\hat{J}}$ is generated by taking products of the
relevant binary codeword sequence elements in the Ising representation 
${\hat{J}_\mu}=\prod_{i \in \mathcal{L}(\mu)} \xi_i$.
Here,
we denote the set of codeword indexes $i$ that participate in the source
index $\mu$ by $\mathcal{L}(\mu)=\{i|a_{\mu i}=1\}$ with $A=(a_{\mu i})$.
Therefore, chosen $i$'s correspond to the ones per row, producing an
Ising version of $\boldsymbol{\hat J}$. 
Note that the additive operation in the Boolean representation is
translated into the multiplication in the Ising one.  
Hereafter, we set $J_\mu,\hat{J}_\mu,\xi_i = \pm 1$
while we do not change the notations for simplicity. 
Furthermore, as we use statistical-mechanics techniques, we
consider the source and codeword-sequence dimensionality ($M$ and $N$,
respectively) to be infinite, keeping the rate $R=N/M$ finite. 

To explore the system's capabilities, we examine the Hamiltonian:
\begin{align}
H(\boldsymbol{S}|\boldsymbol{J})=
\sum_{\mu=1}^M
G[\boldsymbol{S}|J_\mu] \ , \label{eq:4}
\end{align}
with
\begin{align}
G[\boldsymbol{S}|J_\mu]
=-J_\mu \prod_{i \in \mathcal{L}(\mu)}S_i
\label{eq:5}
\end{align}
where we have introduced the dynamical variable $S_i$ to find the optimal
value of $\xi_i$, and $G[\boldsymbol{S}|J_\mu]$ denotes the local connectivity
of a random hypergraph neighboring the source bit $J_\mu$.
It is convenient to represent the posterior probability of
codeword $\boldsymbol{S}$ given a source $\boldsymbol{J}$ in the form
\begin{align}
  \mathcal{P}(\boldsymbol{S}|\boldsymbol{J})
  =
  \frac{\exp[-\beta H(\boldsymbol{S}|\boldsymbol{J})]}{Z(\boldsymbol{J})}
\end{align}
with the inverse tempresure $\beta$,
where $Z(\boldsymbol{J})=\mathrm{Tr}_{\boldsymbol{S}} \exp[-\beta H(\boldsymbol{S}|\boldsymbol{J})]$ is the partition function.

We obtain an expression for the free energy per source bit
expressed in terms of the probability distributions $\pi(x)$ and $\hat{\pi}(\hat{x})$:
\begin{align}
  -\beta f
  &=
  \frac{1}{M} \langle \langle \ln Z(\boldsymbol{J}) \rangle \rangle \nonumber\\
  &= \ln \cosh \beta \nonumber\\
  &{\phantom =}+\int \left[ \prod_{l=1}^K \pi(x_l) dx_l \right]
  \left\langle
  \ln
  \left(
  1+\tanh \beta J \prod_{l=1}^K \tanh \beta x_l
  \right)
  \right\rangle_{J} \nonumber\\
  &{\phantom =}-K \int \pi(x) dx \int \hat{\pi}(\hat{x}) d\hat{x} \ 
  \ln(1+\tanh \beta x \ \tanh \beta \hat{x}) \nonumber\\
  &{\phantom =}+\frac{C}{K}\int
  \left[ \prod_{l=1}^C \pi(\hat{x}_l) d\hat{x}_l \right] \ 
  \ln
  \left[
  {\sum_S}\prod_{l=1}^C (1+S\tanh \beta \hat{x}_l)
  \right] \ , \label{eqn:free}
\end{align}
where $\langle \langle \cdots \rangle \rangle$ denotes the average over
quenched randomness. 
The saddle point equations with respect to probability distributions provide
a set of relations between $\pi(x)$ and $\hat{\pi}(\hat{x})$:
\begin{align}
  \pi(x)=&\int \left[ \prod_{l=1}^C \pi(\hat{x}_l) d\hat{x}_l \right]
  \delta \left(x-\sum_{l=1}^{C-1} \hat{x}_l \right) \ , \label{eqn:SPE1} \\
  \hat{\pi}(\hat{x})=&\int \left[ \prod_{l=1}^K \pi(x_l) dx_l \right] 
  \left\langle \delta \left[\hat{x}-\frac{1}{\beta}\tanh^{-1}
  \left(\tanh \beta J \prod_{l=1}^{K-1} \tanh \beta x_l \right)
  \right] \right\rangle_{J} \ .
 \label{eqn:SPE2}
\end{align}
By using the result obtained for the free energy, we can easily perform
further straightforward calculations to find all the other observable
thermodynamical quantities, including internal energy: 
\begin{align}
 e
 &=
 \frac{1}{M}
 \left\langle\left\langle
 \mathrm{Tr}_{\boldsymbol{S}}
 H(\boldsymbol{S}|\boldsymbol{J})
 e^{-\beta H(\boldsymbol{S}|\boldsymbol{J})}
 \right\rangle\right\rangle
 =
 -\frac{1}{M}
 \frac{\partial}{\partial \beta}
 \langle \langle \ln Z(\boldsymbol{J}) \rangle
 \rangle \ , \label{eqn:energy}
\end{align}
which records reproduction errors. 
This set of equations (\ref{eqn:SPE1}) and (\ref{eqn:SPE2}) may be solved numerically for general $\beta$, $K$, and $C$. 
The spin glass solution can be calculated for both the replica symmetric and
the one step RSB ansatz. The former reduces
to the paramagnetic solution ($f_{\mathrm{RS}}=-1$), which is unphysical
for $R < 1$, while the latter yields continueous distributions
$\pi(x)$ and $\hat{\pi}(\hat{x})$ at the freezing point $\beta_g$,
which can be obtained from the root of the
equation enforcing the non-negative replica symmetric entropy.
The Random-Energy-Model limit $K, C \to \infty$ and simple algebra
gives the relation between the rate $R=N/M$ and the distortion $D$ in the form
$R=1-h_2(D)$,
which coincides with the rate-distortion function in the
Shannon's theorem~\cite{proc:murayama-okada:nips}.

We now take the Thouless-Anderson-Palmer approach to build a dynamics model
using the Markov process assumption on prior beliefs,
showing that it is
possible to obtain a closed set of equations for practical encoding.
At this point, we assume a mean field behavior for the dependence of the
dynamical variables $\boldsymbol{S}$ on a certain realization of the
source sequence $\boldsymbol{J}$, i.e., the dependence is factorizable
and might be replaced by a product of mean fields.
Furthermore, we treat the Boltzmann weights
for specific codeword bit $S_i$
are factorizable with respect to the source bit $J_\mu$~\cite{art:kabashima-saad:tap}.
On the other hand, from a physicist point of view,
it is natural to introduce the Markov process assumption on the priors
to find a solution in the spin glass state.
This term can be considered as the prior knowledge at a certain time $t+1$, given the previous one on the variables at $t$.
Hereafter, we introduce a parameter $t=1,2, \cdots$
to represent time evolution.
In this scenario, we can derive a set of consistency equations:
\begin{align}
  \begin{split}
  \mathcal{W}_t(J_\mu|S_i,\{ J_{\mu^\prime \neq \mu} \})
  =&\sum_{S_{i^\prime \neq i}}
  e^{-\beta G_t(\boldsymbol{S}|J_\mu)}
  \prod_{i^\prime \neq i}
  \mathcal{P}_t(S_{i^\prime}|\{ J_{\mu^\prime \neq \mu} \}) \ , 
  \end{split} 
  \label{eq:7} \\
  \begin{split}
  \mathcal{P}_{t+1}(S_i|\{J_{\mu^\prime \neq \mu}\})
  =&a_{\mu i} \mathcal{Q}_{t+1}(S_i)
  \prod_{\mu^\prime \in \mathcal{M}(i) \backslash \mu}
  \mathcal{W}_t
  (J_{\mu^\prime}|S_i,\{ J_{\mu^{\prime \prime} \neq \mu^\prime} \}) \ ,
  \end{split}
  \label{eq:8}
\end{align}
with
\begin{align}
  \mathcal{Q}_{t+1}(S)
  = \langle \exp(\alpha S+\tanh^{-1} \gamma \cdot S S_i)
  \rangle_{\mathcal{P}_t(S_i|\{ J_{\mu^\prime \neq \mu} \})} \label{eq:8.1}
\end{align}
where $a_{\mu i}$ is a normalization factor.
In (\ref{eq:8.1}), we have two parameters to determine;
$\alpha$ denotes the ferromagnetic bias and $\gamma$ introduces the
autocorrelation for sequences.

Here,
we introduce another set $\mathcal{M}(i)$ such that it defines the set of
source indexes linked to the codeword index $i$.
Equation (\ref{eq:7}) evaluates the average influence of the newly added
parity bit $J_{\mu}$ to $S_i$,
when $\{ S_{i^\prime}|i^\prime \in \mathcal{L}(\mu) \backslash i \}$
obeys a posterior distribution, which should be determined
by the rest of data set
$\{ J_{\mu^\prime}| \mu^\prime \in \mathcal{M}(i)\backslash \mu \}$.
This calculation corresponds to the cavity method in the conventional
framework, representing the effective Boltsmann weight
$\mathcal{W}_t(J_\mu|S_i(t),\{ J_{\mu^\prime \neq \mu} \})$
produced by $J_{\mu}$,
in which the self-induced contributions are eliminated by assuming the tree description for loopy interactions~\cite{book:mezard-parisi-virasoro}.
On the other hand,
equation (\ref{eq:8}) indicates the stack of the cavity fields determines the
posterior distribution $\mathcal{P}_t(S_i|\{J_{\mu^\prime \neq \mu}\})$.

\begin{figure}
\begin{center}
\includegraphics[scale=0.4,angle=0]{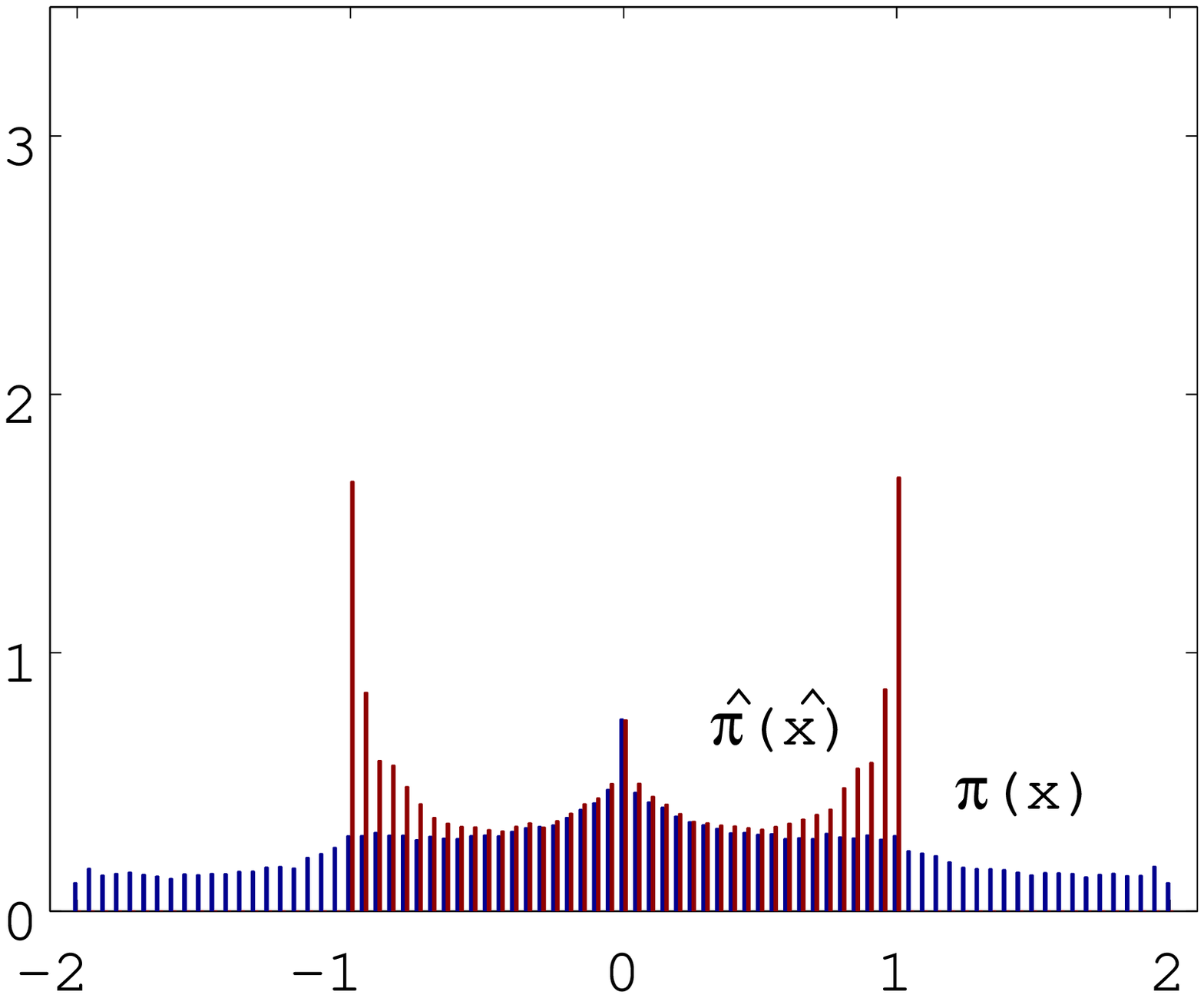}
\includegraphics[scale=0.4,angle=0]{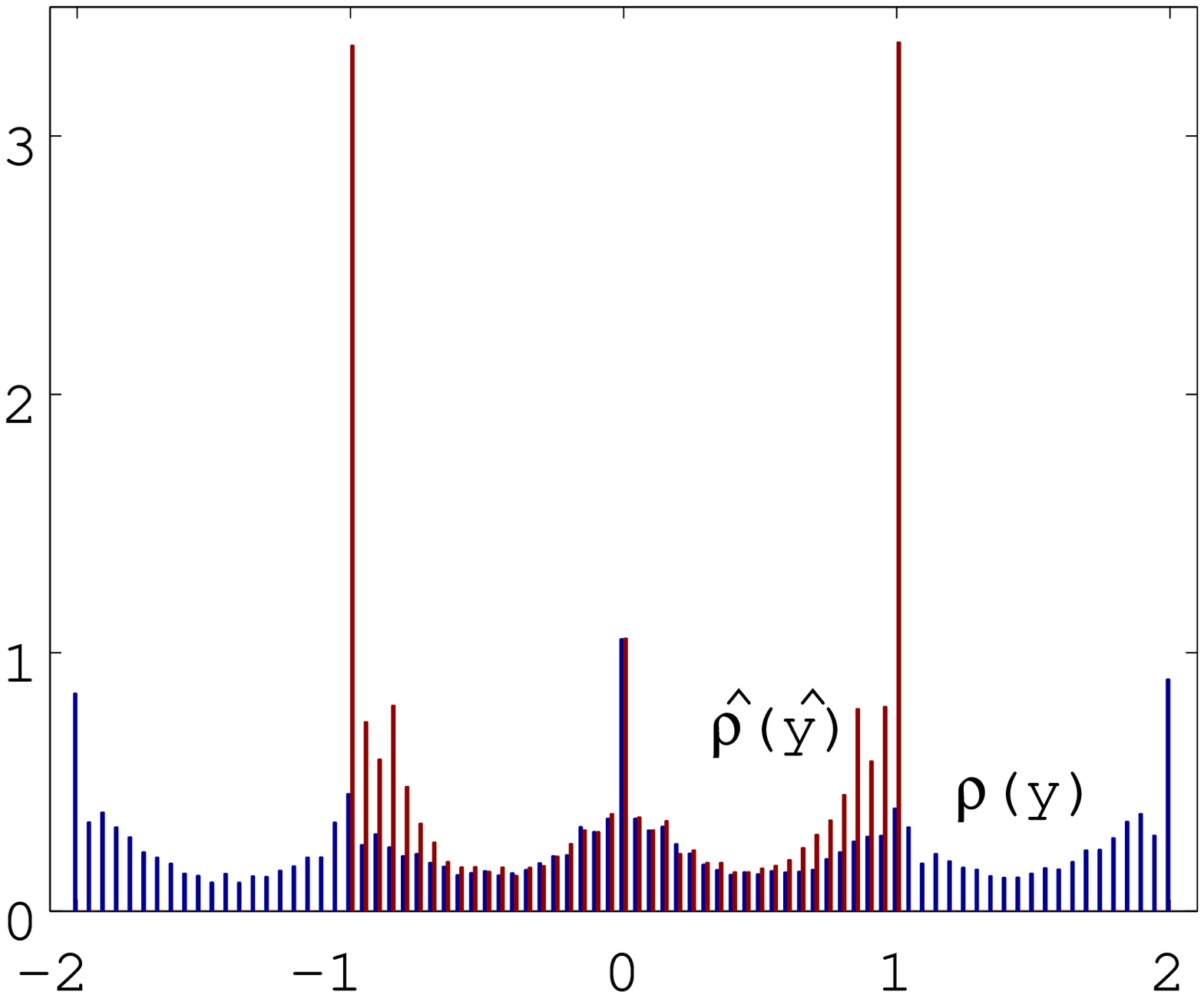}
\end{center}
\caption{
Snapshots of probability distributions for $K=2$, $L=3$ and $\beta_g=2.35$.
(LEFT) Stable solution of (\ref{eqn:SPE1}) and (\ref{eqn:SPE2})
is calculated by Monte Carlo methods.
We use $10^5$ bin models to approximate the probability 
distributions $\pi(x)$ and $\hat{\pi}(\hat{x})$, 
starting from various initial conditions.
(RIGHT) Fixed-point condition for the density evolution of (\ref{eq:10})
and (\ref{eq:11}) is represented in terms of the probability distributions
in the same bin model.
We use the relation $m_{\mu i}=\tanh \beta y$ and
$\hat{m}_{\mu i}=\tanh \beta \hat{y}$,
where the variables are assumed to be generated from common densities
$\rho$ and $\hat{\rho}$, respectively.
}
\label{fig:1}
\end{figure}

In this formula,
the approximated marginal posterior will be
\begin{align}
  \mathcal{P}_{t+1}(S_i|\boldsymbol{J})
  =a_i \mathcal{Q}_{t+1}(S_i)
  \prod_{\mu \in \mathcal{M}(i)}
  \mathcal{W}_t
  (J_\mu|S_i,\{ J_{\mu^\prime \neq \mu} \}) \ ,
\end{align}
taking the full set of the cavity fields,
determined self-consistently by (\ref{eq:7}) and (\ref{eq:8}),
into account, where $a_i$ is a normalization factor again.

Next, we present more convenient form of the above equations.
The conditional probability 
$\mathcal{P}_t(J_\mu|S_i, \{J_{\mu^\prime \neq \mu}\})$
is a normalized effective Boltzmann weight:
\begin{align}
  \mathcal{P}_t(J_\mu|S_i, \{J_{\mu^\prime \neq \mu}\})
  &=b_{\mu i}\mathcal{W}_t
  (J_\mu|S_i,\{J_{\mu^\prime \neq \mu}\}) \nonumber \\
  &=b_{\mu i}\sum_{S_{i^\prime \neq i}}
  e^{-\beta G_t(\boldsymbol{S}|J_\mu)}
  \prod_{i^\prime \neq i} \mathcal{P}_t(S_i|\{J_{\mu^\prime \neq \mu}\}) \ ,
\end{align}
where $b_{\mu i}$ is a normalization constant.
This relation is obtained by taking the connection $\mu$ out of the system,
and taking into consideration the dependance of the variables
$\boldsymbol{S}$ on all other connections.

The identity
\begin{align}
  e^{-\beta G_t(\boldsymbol{S}|J_\mu)}
  =
  \frac{1}{2}
  \cosh (\beta J_\mu)
  \cdot
  \left(
    1+\tanh (\beta J_\mu)
    \prod_{i \in \mathcal{L}(\mu)}S_i
  \right)
\end{align}
and simple algebra with respect to the newly-defined variables
$m_{ij}(t)$, $\hat{m}_{ij}(t) \in [-1,+1]$ satisfying the relations:
\begin{align}
  \mathcal{P}_t(S_i|\{J_{\mu^\prime \neq \mu}\})&=\frac{1+m_{\mu i}(t) S_i}{2} \ , \\
  \mathcal{P}_t(J_\mu|S_i,\{J_{\mu^\prime \neq \mu}\})&=\frac{1+\hat{m}_{\mu i}(t) S_i}{2}
\end{align}
give the set of consistency equations in the form
\begin{align}
  \hat{m}_{\mu i}(t+1)&=
  \tanh(\beta J_\mu)
  \prod_{i^\prime \in \mathcal{L}(\mu)\backslash i}
  m_{\mu i^\prime}(t) \ , \label{eq:10} \\
  m_{\mu i}(t+1)&=
  \tanh
  \left(
    \sum_{\mu^\prime \in \mathcal{M}(i)\backslash \mu}
    \tanh^{-1}
    \hat{m}_{\mu^\prime i}(t)+\alpha+\tanh^{-1} \gamma m_i(t)
  \right) \ , \label{eq:11}
\end{align}
with the pseudo-posterior expresion
\begin{align}
  m_i(t)=
  \tanh
  \left(
    \sum_{\mu \in \mathcal{M}(i)}
    \tanh^{-1}
    \hat{m}_{\mu i}(t)+\alpha+\tanh^{-1} \gamma m_i(t)
  \right) \ . \label{eq:12}
\end{align}
The set of equations (\ref{eq:10}) and (\ref{eq:11})
give an iterative algorithm for code generation.
In our dynamics model,
the choice of parameter $\gamma=0$ results in naive TAP equation
without the reaction term~\cite{book:opper-saad}.
Therefore, in this case, the dynamics can be strictly captured using the method
of ``density evolution'' proposed by Richardson and Urbanke in the context of
determining the capacity of low-density parity-check (LDPC) codes under
message-passing decoding~\cite{art:richardson-urbanke}.
Let $\rho(\cdot)$ denote the common density of
$(1/\beta)\tanh^{-1}m_{\mu i}$,
and $\hat{\rho}(\cdot)$ the density of $(1/\beta)\tanh^{-1}\hat{m}_{\mu i}$
respectively, it is easy to see the set of probability distributions
should satisfy the saddle point equations
(\ref{eqn:SPE1}) and (\ref{eqn:SPE2}).
It is quite interesting to find the consistency between the information
theoretic ``density evolution'' technique and the replica theory for
disordered statistical systems.
Therefore the similarity between $\pi$ and $\rho$
(or $\hat{\pi}$ and $\hat{\rho}$) can be considered as an
important measure for good encoding,
giving the design principle for dynamics model~[FIG.~\ref{fig:1}].
Finally, the equation (\ref{eq:12}) provides the Bayes optimal
encoding $\hat{m}_i(t) = \mathrm{sign}(m_i(t))$ in the Ising representation.

\begin{figure}
\begin{center}
\includegraphics[scale=0.5,angle=0]{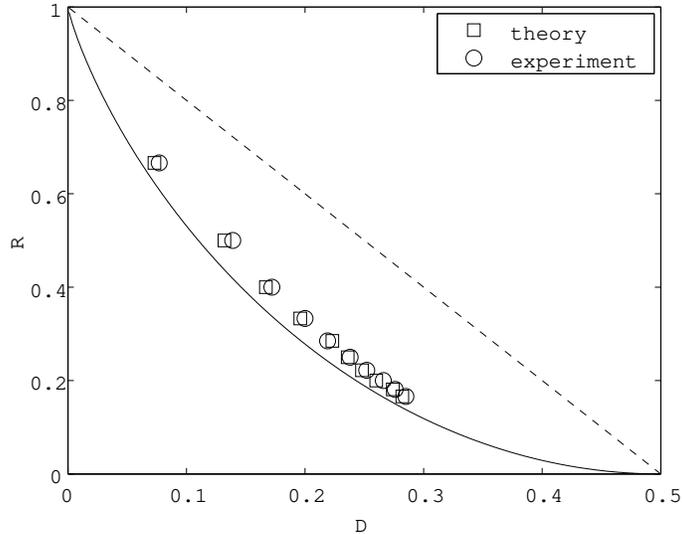}
\end{center}
\caption{Empirical performance:
Numerical experiments show that the algorithm with optimal
$\alpha=0$ and $\beta=\beta_g$
can achieve the bound for sparse construction of the
codes, where $K=2$ and $L=3,4,\cdots,12$.
We choose $\gamma=0.01$ for $C=3$ and $\gamma=0.1$ for the rest.
Solid line denotes the rate-distortion function $R(D)$ for binary sequences
by Shannon, while dashed line can be easily achieved
by universal lossless coding techniques.
($\circ$) Numerical results with the system size $N=20000$, averaged over $10$ trials for each evaluation.
($\square$) Theoretical bound for the sparse construction obtained by the $10^5$ bin model for Monte Carlo sampling in the replica framework.
}
\label{fig:2}
\end{figure}

Practical encoding scheme for this compression model will be as follows.
Given the source sequence $\boldsymbol{J}$, we first translate the
Boolean alphabets into that of Ising ones. Then, for a certain set of control
parameters $\alpha$, $\beta$ and $\gamma$, the equations
(\ref{eq:10}), (\ref{eq:11}) are recursively calculated until they converge
to a certain fix point.
Lastly, according to the equation (\ref{eq:12}), we calculate the
codeword sequence $\boldsymbol{\xi}$ from the Boolean translation
of $\boldsymbol{\hat{m}}=\mathrm{sign}(\boldsymbol{m})$.
Notice that the decoding process will be just a linear mapping.
The most interesting quantity to examine is clearly the minimum typical
distortion for a given compression rate.
Empirical results are shown in FIG.~{\ref{fig:2}},
together with the theoretical evaluation for the code constructions
using the replica method.
We use the optimal inverse temerature $\beta_g$ for code generation,
where per-bit entropy vanishes at the freezing point.
Recent works in the information science reveal that designing the codes
which approach to the Shannon's limit $R(D)$
is quite difficult in the practical sense;
we do not have good coding methods of low complexity,
especially of $\mathcal{O}(N)$.
Our code construction, however, takes only $\mathcal{O}(N)$,
and the performance is surprisingly good.
We believe that the physicist approach can play an important role
in the lossy compression schemes,
as we have already seen in the context of the error-correcting codes.

Future directions of the current research include
utilizing more refined approximation techniques
to find better coding schemes for lossy compression,
as well as the evaluation of the trade-off relation between performance
and computational costs.
These tasks are interesting and challenging.

\begin{acknowledgments}
The author thank Yoshiyuki Kabashima and Jort van Mourik for useful discussions.The author also thank Tadaaki Hosaka for carefully examining the manuscript.
This research was partially supported by the Ministry of Education, Science,
Sports and Culture, Grant-in-Aid for Young Scientists (B),
15760288,
2003.

\end{acknowledgments}

\bibliography{TAP_rev2}

\begin{thebibliography}{10}
\expandafter\ifx\csname natexlab\endcsname\relax\def\natexlab#1{#1}\fi
\expandafter\ifx\csname bibnamefont\endcsname\relax
  \def\bibnamefont#1{#1}\fi
\expandafter\ifx\csname bibfnamefont\endcsname\relax
  \def\bibfnamefont#1{#1}\fi
\expandafter\ifx\csname citenamefont\endcsname\relax
  \def\citenamefont#1{#1}\fi
\expandafter\ifx\csname url\endcsname\relax
  \def\url#1{\texttt{#1}}\fi
\expandafter\ifx\csname urlprefix\endcsname\relax\def\urlprefix{URL }\fi
\providecommand{\bibinfo}[2]{#2}
\providecommand{\eprint}[2][]{\url{#2}}

\bibitem[{\citenamefont{Shannon}(1948)}]{art:shannon}
\bibinfo{author}{\bibfnamefont{C.~E.} \bibnamefont{Shannon}},
  \bibinfo{journal}{Bell Sys. Tech. J.} \textbf{\bibinfo{volume}{27}},
  \bibinfo{pages}{379} (\bibinfo{year}{1948}).

\bibitem[{\citenamefont{Cover and Thomas}(1991)}]{book:cover-thomas}
\bibinfo{author}{\bibfnamefont{T.~M.} \bibnamefont{Cover}} \bibnamefont{and}
  \bibinfo{author}{\bibfnamefont{J.~A.} \bibnamefont{Thomas}},
  \emph{\bibinfo{title}{Elements of Information Theory}}
  (\bibinfo{publisher}{Wiley}, \bibinfo{year}{1991}).

\bibitem[{\citenamefont{Matsunaga and
  Yamamoto}(2002)}]{proc:matsunaga-yamamoto}
\bibinfo{author}{\bibfnamefont{Y.}~\bibnamefont{Matsunaga}} \bibnamefont{and}
  \bibinfo{author}{\bibfnamefont{H.}~\bibnamefont{Yamamoto}}, in
  \emph{\bibinfo{booktitle}{Proceedings 2002 IEEE International Symposium on
  Information Theory}} (\bibinfo{year}{2002}), p. \bibinfo{pages}{461}.

\bibitem[{\citenamefont{Murayama and Okada}(2003)}]{proc:murayama-okada:nips}
\bibinfo{author}{\bibfnamefont{T.}~\bibnamefont{Murayama}} \bibnamefont{and}
  \bibinfo{author}{\bibfnamefont{M.}~\bibnamefont{Okada}}, in
  \emph{\bibinfo{booktitle}{Advances in Neural Information Processing Systems
  15}}, edited by \bibinfo{editor}{\bibfnamefont{S.}~\bibnamefont{Becker}},
  \bibinfo{editor}{\bibfnamefont{S.}~\bibnamefont{Thrun}}, \bibnamefont{and}
  \bibinfo{editor}{\bibfnamefont{K.}~\bibnamefont{Obermayer}}
  (\bibinfo{publisher}{MIT Press}, \bibinfo{year}{2003}).

\bibitem[{\citenamefont{Kabashima and Saad}(1999)}]{art:kabashima-saad:replica}
\bibinfo{author}{\bibfnamefont{Y.}~\bibnamefont{Kabashima}} \bibnamefont{and}
  \bibinfo{author}{\bibfnamefont{D.}~\bibnamefont{Saad}},
  \bibinfo{journal}{Europhys. Lett.} \textbf{\bibinfo{volume}{45}},
  \bibinfo{pages}{97} (\bibinfo{year}{1999}).

\bibitem[{\citenamefont{Sourlas}(1989)}]{art:sourlas:nature}
\bibinfo{author}{\bibfnamefont{N.}~\bibnamefont{Sourlas}},
  \bibinfo{journal}{Nature} \textbf{\bibinfo{volume}{339}},
  \bibinfo{pages}{693} (\bibinfo{year}{1989}).

\bibitem[{\citenamefont{Kabashima and Saad}(1998)}]{art:kabashima-saad:tap}
\bibinfo{author}{\bibfnamefont{Y.}~\bibnamefont{Kabashima}} \bibnamefont{and}
  \bibinfo{author}{\bibfnamefont{D.}~\bibnamefont{Saad}},
  \bibinfo{journal}{Europhys. Lett.} \textbf{\bibinfo{volume}{44}},
  \bibinfo{pages}{668} (\bibinfo{year}{1998}).

\bibitem[{\citenamefont{Mezard et~al.}(1987)\citenamefont{Mezard, Parisi, and
  Virasoro}}]{book:mezard-parisi-virasoro}
\bibinfo{author}{\bibfnamefont{M.}~\bibnamefont{Mezard}},
  \bibinfo{author}{\bibfnamefont{G.}~\bibnamefont{Parisi}}, \bibnamefont{and}
  \bibinfo{author}{\bibfnamefont{M.}~\bibnamefont{Virasoro}},
  \emph{\bibinfo{title}{Spin-Glass Theory and Beyound}}
  (\bibinfo{publisher}{World Scientific}, \bibinfo{year}{1987}).

\bibitem[{\citenamefont{Opper and Saad}(2001)}]{book:opper-saad}
\bibinfo{author}{\bibfnamefont{M.}~\bibnamefont{Opper}} \bibnamefont{and}
  \bibinfo{author}{\bibfnamefont{D.}~\bibnamefont{Saad}},
  \emph{\bibinfo{title}{Advanced Mean Field Methods: Theory and Practice}}
  (\bibinfo{publisher}{MIT Press}, \bibinfo{year}{2001}).

\bibitem[{\citenamefont{Richardson and Urbanke}(2001)}]{art:richardson-urbanke}
\bibinfo{author}{\bibfnamefont{T.~J.} \bibnamefont{Richardson}}
  \bibnamefont{and} \bibinfo{author}{\bibfnamefont{R.~L.}
  \bibnamefont{Urbanke}}, \bibinfo{journal}{IEEE Trans. Inf. Theory}
  \textbf{\bibinfo{volume}{47}}, \bibinfo{pages}{599} (\bibinfo{year}{2001}).

\end{thebibliography}

\end{document}